\PassOptionsToPackage{table}{xcolor}
\documentclass[twocolumn,prd,preprintnumbers,numbers,sort&compress,nofootinbib,showpacs,colorlinks,citecolor=blue,amsmath,amssymb,aps,superscriptaddress]{revtex4-2}

\usepackage{graphicx}
\usepackage{dcolumn}
\usepackage{bm}
\usepackage{xcolor}
\usepackage{lipsum} 
\usepackage{float} 
\usepackage[utf8]{inputenc}

\usepackage{CJKutf8} 

\newcommand{\be}{\begin{equation}}
\newcommand{\ee}{\end{equation}}
\newcommand{\bw}{\begin{widetext}}
\newcommand{\ew}{\end{widetext}}
\newcommand{\ba}{\begin{aligned}}
\newcommand{\ea}{\end{aligned}}
\newcommand{\bes}{\begin{equation*}}
\newcommand{\ees}{\end{equation*}}
\newcommand{\bea}{\begin{eqnarray}}
\newcommand{\eea}{\end{eqnarray}}
\newcommand{\dd}{\text{d}}

\newcommand{\hmC}{\hat{\mathcal{C}}}

\newcommand{\fR}{\mathfrak{R}}
\newcommand{\hfR}{\hat{\mathfrak{R}}}
\newcommand{\tFD}{\mathcal{T}_\mathrm{FD}}

\newcommand{\beq}{\begin{equation}}
\newcommand{\eeq}{\end{equation}}
\newcommand{\exclude}[1]{}

\newcommand{\pasj}{Publications of the Astronomical Society of Japan}
\newcommand{\ssr}{Space Science Reviews}

\newcommand{\aap}{Astronomy \& Astrophysics}

\newcommand{\mnras}{MNRAS}
\newcommand{\apjs}{ApJ\ Supp.}

\newcommand{\physrep}{Physics Reports}

\definecolor{Orange}{rgb}{1.0,0.5,0.15}
\definecolor{Blue}{rgb}{0,0.08,0.65}
\definecolor{Red}{rgb}{0.65,0.08,0.05}
\definecolor{Green}{rgb}{0.15,0.45,0.25}
\definecolor{Pink}{rgb}{1.0,0.05,0.5}
\definecolor{bubbles}{rgb}{0.91, 1.0, 1.0}
\definecolor{aquamarine}{rgb}{0.5, 1.0, 0.83}
\definecolor{bubblegum}{rgb}{0.99, 0.76, 0.8}
\definecolor{bluebell}{rgb}{0.74, 0.74, 0.92}
\definecolor{dollarbill}{rgb}{0.72, 0.93, 0.6}
\definecolor{cred}{RGB}{238,28,37}

\newcommand{\ubc}{Department of Physics \& Astronomy, University of British Columbia, 6224 Agricultural Road, Vancouver, BC V6T 1Z1, Canada}

\newcommand{\bochum}{Astronomisches Institut, Ruhr-Universität Bochum, Universitätsstr. 150, 44801, Bochum, Germany}

\newcommand{\cea}{Universit\'e Paris-Saclay, IPHT, DRF-INP, UMR 3681, CEA, Orme des Merisiers Bat 774, 91191 Gif-sur-Yvette, France}

\newcommand{\cgpqi}{Center for Gravitational Physics and Quantum Information, Yukawa Institute for Theoretical Physics, Kyoto University, Kyoto 606-8502, Japan}

\newcommand{\haa}{National Research Council Herzberg Astronomy and Astrophysics, 5071 West Saanich Road, Victoria, B.C., V8Z 6M7, Canada}
\newcommand{\ubccs}{Department of Computer Science, University of British Columbia, 2366 Main Mall, Vancouver, BC V6T 1Z4, Canada}

\newcounter{FFcounter}

\newcounter{LVWcounter}

\newcounter{SGcounter}

\newcounter{RDcounter}

\usepackage{hyperref}

\begin{document}

\preprint{APS/123-QED}

\title{Mitigating Nonlinear Systematics in Weak Lensing Surveys II: Stability and Diagnostics with Intrinsic Alignment}

\author{Shiming Gu (\begin{CJK}{UTF8}{gbsn}顾时铭\end{CJK})}
 \email{gsm@phas.ubc.ca}
\affiliation{\ubc}
\affiliation{\bochum}

\author{Ludovic van Waerbeke}%
 \email{waerbeke@phas.ubc.ca}
\affiliation{\ubc}

\author{Francis Bernardeau}
      \affiliation{\cea}
      \affiliation{\cgpqi}

\author{S\'ebastien Fabbro}
\affiliation{\haa}
\affiliation{\ubccs}

\date{\today}

\begin{abstract}
The Bernardeau–Nishimichi–Taruya (BNT) transform provides a powerful framework for analysing tomographic cosmic shear data by improving the localization of shear correlations in physical scale. It operates by performing a linear combination of the shear data vector in $\ell$-space, yielding a transformed vector that is better localized in both redshift and $k$-space. BNT is particularly useful for estimating cosmological parameters while minimizing the impact of poorly understood nonlinear physics, without discarding large amounts of information as is typically done with simple scale cuts. In our previous work, we showed that BNT outperforms traditional weak-lensing analyses; however, that study did not include intrinsic alignments (IA). In the present work, we assess the robustness of our BNT-based $k$-cut framework in the presence of realistic IA models. We consider two cases: (i) when the assumed IA model used in sampling is close to, but not identical to, the true one, and (ii) when the assumed IA model is significantly {biased compared to the true one}. In the first case, the $k$-cut framework yields precise and unbiased $S_8$ constraints even with limited knowledge of large-scale modes. {Using Euclid-like mock data and a stringent $k$-cut of $k\le 0.1\,{\rm Mpc^{-1}}$ for all tomographic bins, we found that BNT can constrain $S_8$ with a precision better than 2\% while non-BNT has lost all constraining power.} In the second case, the BNT transform serves as a powerful diagnostic tool, revealing internal inconsistencies in $k$-space and redshift-space both exceeding 5$\sigma$ when the functional form of the sampling and fiducial IA models differ fundamentally.
\end{abstract}

\maketitle

\section{Introduction}

Weak gravitational lensing as a probe of dark matter is a core science objective of Stage IV imaging surveys (e.g., Euclid \cite{Euclid}, Rubin \cite{LSST}, Xuntian \cite{CSST}). The substantial gain in statistical power provides a powerful means to identify and reduce residual systematics, whether instrumental (e.g., Point Spread Function characterization and correction) or theoretical (e.g., modeling of the nonlinear matter power spectrum). At the precision expected from Stage IV surveys, even subtle biases arising from incomplete models, instrumental artifacts, or intrinsic astrophysical correlations could dominate the error budget and bias cosmological inference. Mitigating these biases is now a defining challenge for weak-lensing cosmology.

Weak gravitational lensing is fundamentally a two-dimensional probe: for a given source redshift, it measures anisotropies in the mass distribution projected along the line of sight from the sources to the observer ({for a review on WL, see \citep{2015RPPh...78h6901K}}). This makes it challenging to identify biases in the nonlinear three-dimensional modeling of the matter power spectrum, because only two-dimensional projections are observed. Consequently, the usual approach is to apply scale cuts to the WL data vector, at the cost of significant information loss. {Recent stage III surveys have shown quantitatively how scale cuts limit the precision of cosmological inference \citep{Troxel2018DESy1,2019PASJ...71...43H,2021A&A...645A.104A}}.

However, \cite{Bernardeau2014BNT} showed that a certain linear combination of weak-lensing observables can recover some of the three-dimensional information. This is possible because, for a lens at a given redshift, the redshift scaling of the WL effect is purely geometric and independent of the matter distribution—a property that underlies nulling techniques \cite{2008A&A...488..829J,Touzeau2025BNT2}. The specific linear combination that achieves this, developed in \cite{Bernardeau2014BNT}, is called the Bernardeau–Nishimichi–Taruya (BNT) transform. In our previous work \cite{Gu2025BNT1} (Paper I), we adapted and tested this transform as a targeted solution for mitigating nonlinear systematics.

We showed that a BNT-transformed data vector, combined with a well-defined angular scale filtering ($\ell$-cut in harmonic space), can effectively operate as a $k$-cut, enabling more robust handling of uncertainties in the nonlinear modeling of the three-dimensional matter power spectrum. However, in Paper I, we did not account for intrinsic alignment (IA), a known contaminant of the WL signal that can mimic lensing by correlating galaxy shapes over large distances ({for reviews on IA, see \citep{2015PhR...558....1T,2015SSRv..193....1J}}). In this paper, we revisit a number of results obtained in paper I and reevaluate the performance of the BNT transform when IA is taken into account.

The paper is organised as follows. In Section II, we review the modelling of cosmic shear and intrinsic alignment signals, along with their corresponding BNT transforms. In Section III, we describe our methodology for calculating physical and redshift scale cuts using BNT, and outline our posterior sampling procedure. In Section IV, we present forecasts of parameter constraints using both BNT and non-BNT approaches, applied to Euclid-like mock data with non-zero intrinsic alignment, and demonstrate the tests and consistency checks introduced above. Finally, in Section V, we summarise our findings and offer concluding remarks.

\section{Theory}

\subsection{Cosmic Shear}

We consider a cosmological survey with measured photometric redshifts for galaxies, enabling tomographic binning with source-galaxy distributions $n_i(\chi)$ for the $i$th bin, where $\chi$ is the radial comoving distance. The two-dimensional cosmic-shear power spectrum $C^{i,j}(\ell)$ between tomographic bins $(i,j)$ is given by the line-of-sight projection of the non-linear three-dimensional matter power spectrum $P_{\rm nl}(k;z)$, as follows \cite{Limber1954Projection}:

\be
C^{i,j}(\ell) = \int\dfrac{d\chi}{\chi^2} W^i_\gamma(\chi)W^j_\gamma(\chi)P_{\rm nl}\left(k=\dfrac{\ell+1/2}{\chi};z(\chi)\right),
\label{Cell_original}
\ee
where $\chi$ is the radial comoving distance at redshift $z$, and $W_\gamma^i(\chi)$ is the lensing kernel computed by the source redshift distribution of the $i^{th}$ tomographic bin:
\be
W_\gamma^i(\chi) = \dfrac{3\Omega_mH_0^2}{2c^2}\int\dd \chi' \dfrac{n_i(\chi')}{a(\chi)}\dfrac{(\chi'-\chi)\chi}{\chi'},
\label{Wdef}
\ee
where $a(\chi)$ is the scale factor, $H_0$ is the Hubble constant today and $\Omega_m$ the total matter density today. We have assumed a flat Universe with zero spatial curvature, $K=0$. For non-flat geometry, the radial comoving distance $\chi$ has to be replaced by the comoving angular diameter distance $f_K(\chi)$ in Eq. \ref{Wdef}:

$$
f_K(\chi)=
\begin{cases}
\dfrac{1}{\sqrt{K}}\;\sin\!\big(\sqrt{K}\,\chi\big), & K>0\ \text{(closed)},\\[6pt]
\chi, & K=0\ \text{(flat)},\\[6pt]
\dfrac{1}{\sqrt{-K}}\;\sinh\!\big(\sqrt{-K}\,\chi\big), & K<0\ \text{(open)}.
\end{cases}
$$

\subsection{Intrinsic Alignment}

In addition to the coherent distortions induced by gravitational lensing described by Eq.~\ref{Cell_original}, intrinsic alignments (IA) generate additional angular correlations among galaxy shapes—both within a given tomographic bin, through local tidal interactions, and across widely separated bins, via long-range correlations. {The relevant IA contributions are the intrinsic-intrinsic "II", which corresponds to the correlated galaxy shapes caused by their physical proximity, and the gravitational-intrinsic "GI", which corresponds to the correlation between the lensed galaxy shape caused by a large-scale structure and the intrinsic galaxy shape located in that structure. We also define the "GG" term, which corresponds to the lensing signal (gravitational-gravitational) without IA contribution. The resulting observed correlation between tomographic bins $i$ and $j$ can be written as:

\be
C^{i,j} \equiv C^{i,j}_{\rm GG} + C^{i,j}_{\rm GI} + C^{i,j}_{\rm IG} + C^{i,j}_{\rm II},
\label{eq:C_ij_all_def}
\ee
where the angular power spectra can be expressed as line-of-sight integrals of the three-dimensional power spectra $P_\mathrm{GG}$, $P_\mathrm{GI}$ and $P_\mathrm{II}$:
\bea
C^{i,j}_{\rm GG}(\ell) &=& \int\dfrac{d\chi}{\chi^2} W^i_\gamma(\chi)W^j_\gamma(\chi)P_\mathrm{GG}(k;z) \label{eqn:Cgg}\\
C^{i,j}_{\rm GI}(\ell) &=& \int\dfrac{d\chi}{\chi^2} W^i_\gamma(\chi)W^j_N(\chi)P_\mathrm{GI}(k;z) \label{eqn:Cgi} \\
C^{i,j}_{\rm IG}(\ell) &=& \int\dfrac{d\chi}{\chi^2} W^i_N(\chi)W^j_\gamma(\chi)P_\mathrm{GI}(k;z) \label{eqn:Cig} \\
C^{i,j}_{\rm II}(\ell) &=& \int\dfrac{d\chi}{\chi^2} W^i_N(\chi)W^j_N(\chi)P_\mathrm{II}(k;z) \label{eqn:Cii}
\eea
where $W^i_N(\chi)$ is the number density projection kernel:
\be
W^i_N(\chi) = b^i\frac{H(\chi)}{c} \, n_i(z(\chi)),
\ee
with $b^i$ being the linear galaxy bias of i$^{th}$ tomographic bin and $H(\chi)$ is the Hubble constant at $\chi$. In this paper, in order to keep the focus on the intrinsic alignment, we set $b^i = 1$.}

These intrinsic correlations can mimic or bias the lensing signal, making IA a significant astrophysical systematic in weak-lensing analyses \cite{Hirata2004IA}. {The GI, IG and II terms explicitly depend on the IA model.}

One of the simplest IA models is the Non-linear Alignment (NLA) model, which extends the linear alignment framework \cite{Catelan2001IA,Hirata2004IA} by replacing the linear matter power spectrum with its nonlinear counterpart \cite{Bridle2007NLA}. It assumes that intrinsic alignments arise from a tidal interaction that linearly couples galaxy shapes to the local large-scale gravitational field, while nonlinear clustering effects are incorporated through the use of $P_{\mathrm{nl}}(k)$. The NLA model can be described by a redshift-dependent linear bias $C_1(z)$ applied to the nonlinear matter power spectrum:
\bea
P_\mathrm{GG}(k, z) &=& P_{\mathrm{nl}}(k; z) \nonumber\\
P_\mathrm{GI}(k, z) &=& C_1(z) \, P_{\mathrm{nl}}(k; z) \nonumber\\
P_\mathrm{II}(k, z) &=& C_1^2(z) \, P_{\mathrm{nl}}(k, z)
\label{eq:pofkz}
\eea
where the tidal alignment bias $C_1(z)$ is always parameterised as a redshift-dependent function involving a global IA scaling amplitude $A_\mathrm{TA}$ and a power-law evolution index $\eta_\mathrm{TA}$:
\be \label{eqn:c1}
C_1(z) \equiv -A_\mathrm{TA} \, \tilde{C}_1 \, \frac{\rho_\mathrm{crit} \, \Omega_m}{D(z)} \left( \frac{1+z}{1+z_0} \right)^{\eta_\mathrm{TA}}
\ee
where $\rho_\mathrm{crit}$ is the critical density of the universe, $D(z)$ is the structure growth function, $z_0$ is a pivot redshift fixed to $z_0 = 0.62$ \cite{Troxel2018DESy1}, and $\tilde{C}_1 = 5\times10^{-14}\,h^{-2}\mathrm{M}_\odot^{-1}\mathrm{Mpc}^3$ is a normalisation constant.

There is more than one approach for addressing the redshift dependence of the intrinsic alignment. A commonly used alternative to the $\eta_\mathrm{TA}$ parametrization involves the $B_\mathrm{TA}$ parameter in a so-called NLA-z model \cite{Fortuna2021IALRG}:
\be \label{eq:nlaz}
C_1(z) = -\tilde{C}_1 \, \frac{\rho_\mathrm{crit} \, \Omega_m}{D(z)}\left[A_\mathrm{TA} + B_\mathrm{TA}\left(\frac{\langle a(z) \rangle_i}{a_\mathrm{piv}} - 1\right)\right],
\ee
where $\langle a(z) \rangle_i$ is the $n(z)$-weighted average scale factor in tomographic bin $i$, and $a_\mathrm{piv} = 0.769$ is the pivotal scale factor.
Note that, unlike the $\eta_\mathrm{TA}$ parameterisation of NLA model (Eq.\ref{eqn:c1}), this NLA-z model can change sign as a function of redshift.

Another frequently-used model is the Tidal Alignment and Tidal Torquing (TATT) model \cite{Blazek2019TATT}, which extends beyond the NLA framework by incorporating both linear (Tidal Alignment, TA) and quadratic (Tidal Torquing, TT) responses of galaxy shapes to the tidal field. Developed to capture intrinsic alignments of both early- and late-type galaxies within a unified framework, the TATT model introduces additional degrees of freedom to describe more complex IA behaviour. However, even if TATT offers greater physical flexibility, its additional parameters are not yet tightly constrained by current data, leading to substantial degeneracies and potential biases when applied to lensing surveys \cite{Leonard2024TATT}.

In the TATT model, the intrinsic shape of a galaxy can be  written as an expansion of the tidal field tensor $s_{ij}$ where \cite{Blazek2019TATT}:

\be
\gamma_{ij}^\mathrm{IA} = C_\mathrm{1}s_{ij} + C_{1\delta}(\delta s_{ij}) + C_2\left[\sum_{k=0}^{2}s_{ik}s_{kj} - \frac{1}{3}\delta_{ij}s^2\right] + ...
\label{gamma_TATT}
\ee
where $\delta$ is the matter overdensity, $\delta_{ij}$ is the kronecker delta, and tidal field tensor $s_{ij}$ is:
\be
s_{ij}({\bm k}) = \left(\frac{k_ik_j}{k^2}-\frac{1}{3}\delta_{ij}\right)\delta({\bm k})
\ee
with $k^2 = k_1^2 + k_2^2 + k_3^2$, and $k$ is the modulus of the 3-dimensional wavevector $\bm k$. For $C_{1\delta}$ and $C_2$, a similar description like Eq. \ref{eqn:c1} can be used with some minor modifications:
\bea
C_{1\delta}(z) &=& b_\mathrm{TA}C_1(z) \nonumber\\
C_2(z) &=& -5A_\mathrm{TT} \, \tilde{C}_1 \, \frac{\rho_\mathrm{crit} \, \Omega_m}{D^2(z)} \left( \frac{1+z}{1+z_0} \right)^{\eta_\mathrm{TT}}
\label{eq:TATTparams}
\eea
where $A_\mathrm{TT}$ is the Tidal Torquing scaling amplitude and $b_\mathrm{TA} > 0$ is the density weighting bias. To avoid introducing too many equations, please refer to the Section III. of \cite{Blazek2019TATT} for the perturbation theory-based prescription to compute $P_\mathrm{GI}(k)$ and $P_\mathrm{II}(k)$ under the TATT model. It is worth mentioning that, for the TATT model, Eq. \ref{eq:pofkz} cannot be applied. This is a consequence of Eq. \ref{gamma_TATT} which implies that the redshift and $k$ dependence can no longer be factorized. For an effective description of the intrinsic alignment amplitude, compared to the standard NLA model (Eq. \ref{eqn:c1}), we also define an \textit{effective tidal alignment amplitude} $A_\mathrm{TA,eff}$:
\be\label{eqn:A_ta_eff}
A_\mathrm{TA,eff}(k,z) \equiv \dfrac{D(z)}{\tilde{C}_1\rho_\mathrm{crit}\Omega_m}\sqrt{\dfrac{P_\mathrm{II}(k,z)}{P_\mathrm{GG}(k,z)}}
\ee

The effective amplitude $A_\mathrm{TA,eff}$ can be calculated for any IA model used in this work (Eq.\ref{eqn:c1} for NLA, Eq.\ref{eq:nlaz} for NLAz and Eq.\ref{eq:TATTparams} for TATT), and will be used in Section IV to compare the model between each other.

\subsection{BNT Transform}
\label{subsec:IIA}

The BNT transform is a linear combination of the data vector $C^{i,j}(\ell)$ into a new one $\hat C^{a,b}(\ell)$, where $a$ and $b$ represent a linear combination of the original tomographic redshift bins $i$ and $j$. Throughout the paper, we will keep this `hat' notation $\hat{ }$ for all BNT transformed quantities, and reserve the use of $(i,j)$ for the original, noBNT tomographic bins and $(a,b)$ for the BNT tomographic bins.

\begin{figure}[htbp]
\centering
\includegraphics[width=0.45\textwidth, trim = 0.1cm 4.35cm 0.1cm 0cm, clip]{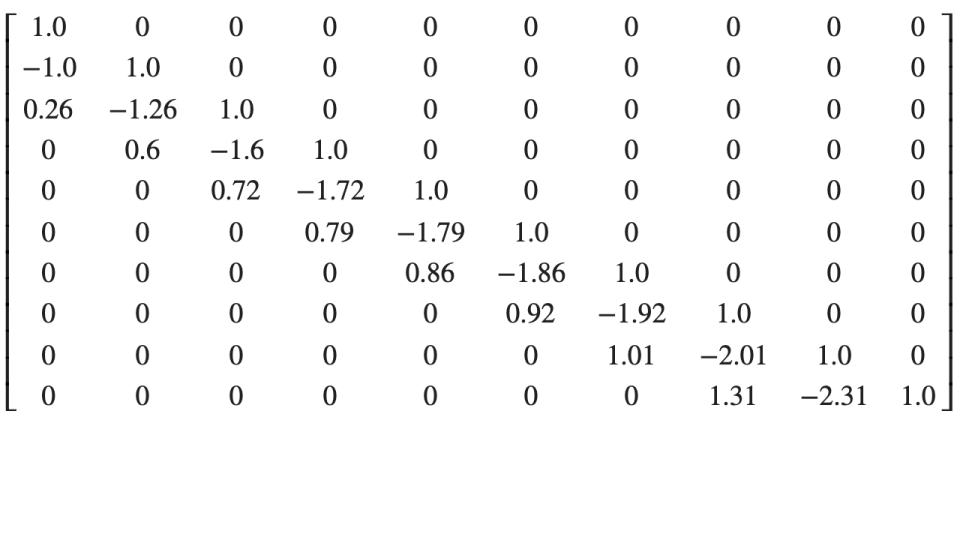}
\caption{
The BNT transform matrix $p^a_i$ calculated with our fiducial cosmology and redshift distribution.
} \label{fig:BNT_pai}
\end{figure}

The BNT transform matrix $p^a_i$ is defined as the following. Two normalisation numbers, $n_i^{0}$ and $n_i^{1}$, are first calculated for each tomographic bin $i$:
\bea
n_i^{0} &=& \int\dd\chi \; n_i(\chi) \\
n_i^{1} &=& \int\dd\chi \; \dfrac{n_i(\chi)}{\chi}
\eea
The transformation matrix elements $p^a_i$ are then constructed from the following algebraic equations:
\bea
&\sum_{i=a-2}^a& p^a_{i} n_i^{0} = 0 \nonumber \\
&\sum_{i=a-2}^a& p^a_{i} n_i^{1} = 0,
\label{p_a_i_elements}
\eea
with $p^a_i = 0$ when $i \notin \{a-2, a-1, a\}$, $p_1^1 = p_2^2 = 1$, and $p_2^1 = -1$. The matrix $p^a_i$ is hence a $n_{\rm T}\times n_{\rm T}$ square matrix, where $n_{\rm T}$ is the number of tomographic bins, with non-zero elements only in a diagonal band of a width of three. Fig. \ref{fig:BNT_pai} shows the BNT transformation matrix for our fiducial cosmology: the columns represent the original tomographic bin entries and the rows represent the BNT tomographic bin entries. One can see that for all rows, except the first, the sum of the matrix elements is always zero. This is a feature of the transform construction, which is built from the algebraic nulling constraints in Eq. \ref{p_a_i_elements}.

Unlike Paper I \cite{Gu2025BNT1}, where intrinsic alignment was not included in the BNT modelling, we now apply the BNT matrix to the full data vector defined in Eq. \ref{eq:C_ij_all_def}, including the intrinsic-alignment terms. This is the only consistent approach, as the observer cannot separate the lensing component $C_{\gamma^{i},\gamma^{j}}$ from the other contributions:
\be
\hat C^{a,b}(\ell) = p^a_ip^b_jC^{i,j}(\ell)
\label{Cell_BNT}
\ee
It is worth mentioning that, for the NLA model, combining Eq. \ref{eq:pofkz} and Eqs. \ref{eqn:Cgg}-\ref{eqn:Cii} leads to a compact form for the BNT data vector $\hat C^{a,b}(\ell)$:

\begin{equation}
    \hat C^{a,b}(\ell)=\int\dfrac{d\chi}{\chi^2} \hat W^a_{\rm eff}(\chi)\hat W^b_{\rm eff}(\chi)P_\mathrm{GG}(k;z),
    \label{eq:hatCell_NLA}
\end{equation}
where we have defined:
\bea
\hat{W}^a_\mathrm{eff}(\chi)\hat{W}^b_\mathrm{eff}(\chi) &\equiv& \hat{W}_\gamma^a(\chi)\hat{W}_\gamma^b(\chi) + \hat{W}_{\gamma}^a(\chi)\hat{W}_{I}^b(\chi) \nonumber \\
&+& \hat{W}_{I}^a(\chi)\hat{W}_{\gamma}^b(\chi) + \hat{W}^a_{I}(\chi)\hat{W}^b_{I}(\chi),
\eea
with:
\bea
\hat W_\gamma^a(\chi) &\equiv& \sum_{i=1}^{n_{\rm T}} p^a_i W_\gamma^i(\chi)\nonumber\\
\hat{W}^a_{I}(\chi) &\equiv& \sum_{i=1}^{n_T}p^a_i\left[C^i_1(\chi)W_N^i(\chi)\right].
\eea
With this definition, $\hat W_\gamma^a(\chi)$ is the BNT-transformed lensing kernel and $\hat{W}^a_{I}(\chi)$ is the BNT-transformed intrinsic-alignment kernel. It is important to note that, by construction, the BNT lensing kernel is localized in redshift, whereas this is not necessarily true for the BNT intrinsic-alignment kernel. The remainder of this work quantifies how intrinsic alignments affect the ability of BNT to separate angular from physical scales in the lensing term. For TATT in particular, the factorization in Eq. \ref{eq:hatCell_NLA} is not applicable due to the $k$ dependence in the IA kernel.

\begin{figure}[htbp]
\centering
\includegraphics[width=0.5\textwidth, trim = 0.6cm 0.2cm 1.0cm 0.2cm, clip]{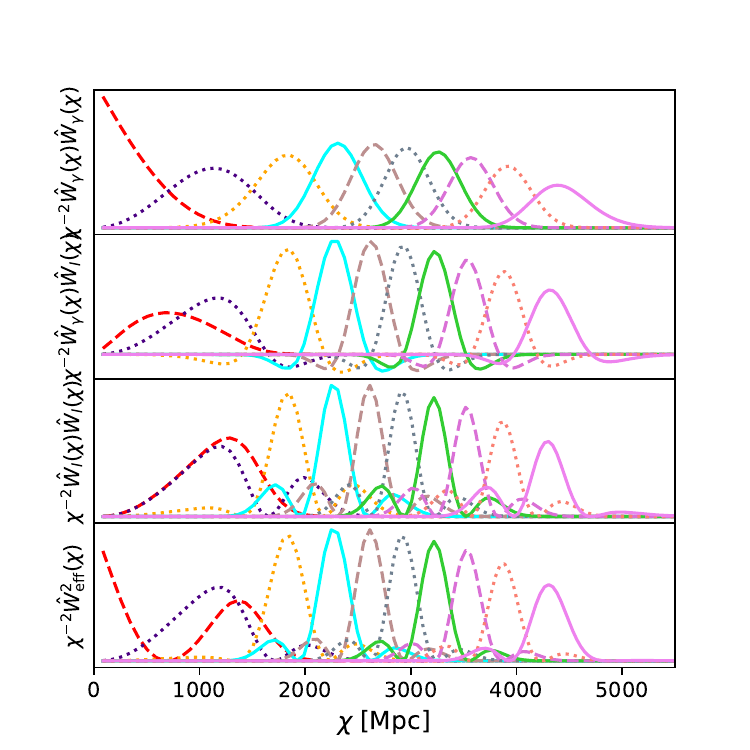}
\caption{Panel 1 of 4: BNT Lensing projection weight $\hat W_\gamma^2(\chi)/\chi^2$ for the shear-shear term. The original BNT projection kernel.
Panel 2 of 4: BNT Lensing projection weight $\hat{W}_{\gamma}(\chi)\hat{W}_{I}(\chi)/\chi^2$ for the shear-shape term with $W_{I}(\chi) = C_1(\chi)W_{N}(\chi)$ under the NLA model with $A_\mathrm{TA} = 1.71$.
Panel 3 of 4: BNT Lensing projection weight $\hat{W}_{I}^2(\chi)/\chi^2$ for the shape-shape term with $W_{I}(\chi) = C_1(\chi)W_{N}(\chi)$ under the NLA model with $A_\mathrm{TA} = 1.71$.
Panel 4 of 4: The total projection weight $W^2_\mathrm{eff}(\chi) = \hat{W}_\gamma^2(\chi) + 2\hat{W}_{\gamma}(\chi)\hat{W}_{I}(\chi) + \hat{W}_{I}^2(\chi)$ for the shear and shape included total $C_\ell$.
} \label{fig:nz}
\end{figure}

In the rest of this work, we will use an Euclid-like mock with ten redshift bins weak lensing tomography \cite{Euclid2020Forecast}. In (mock-) observational practices, galaxies are binned by their photometric redshift estimates. Because photometric redshifts have finite uncertainty, the true redshift distribution $n(z)$ of each bin is broadened and overlaps with its neighboring bins, even if the original photometric binning is non-overlapping. We can compute the kernels directly from the redshift distributions, and Figure \ref{fig:nz} displays the three possible kernel combinations as functions of $\chi$, alongside the effective kernel $\chi^{-2}\hat{W}^2_\mathrm{eff}(\chi)$. The BNT intrinsic-alignment kernel is less localized in redshift than the BNT lensing kernel, yet it remains more localized than the original noBNT lensing kernel.

\section{Methodology}

\subsection{Scale Cuts}
\label{subsec:IIIA}

\begin{figure*}[htbp]
\centering
\includegraphics[width=0.95\textwidth, trim = 6cm 6cm 6cm 6cm, clip]{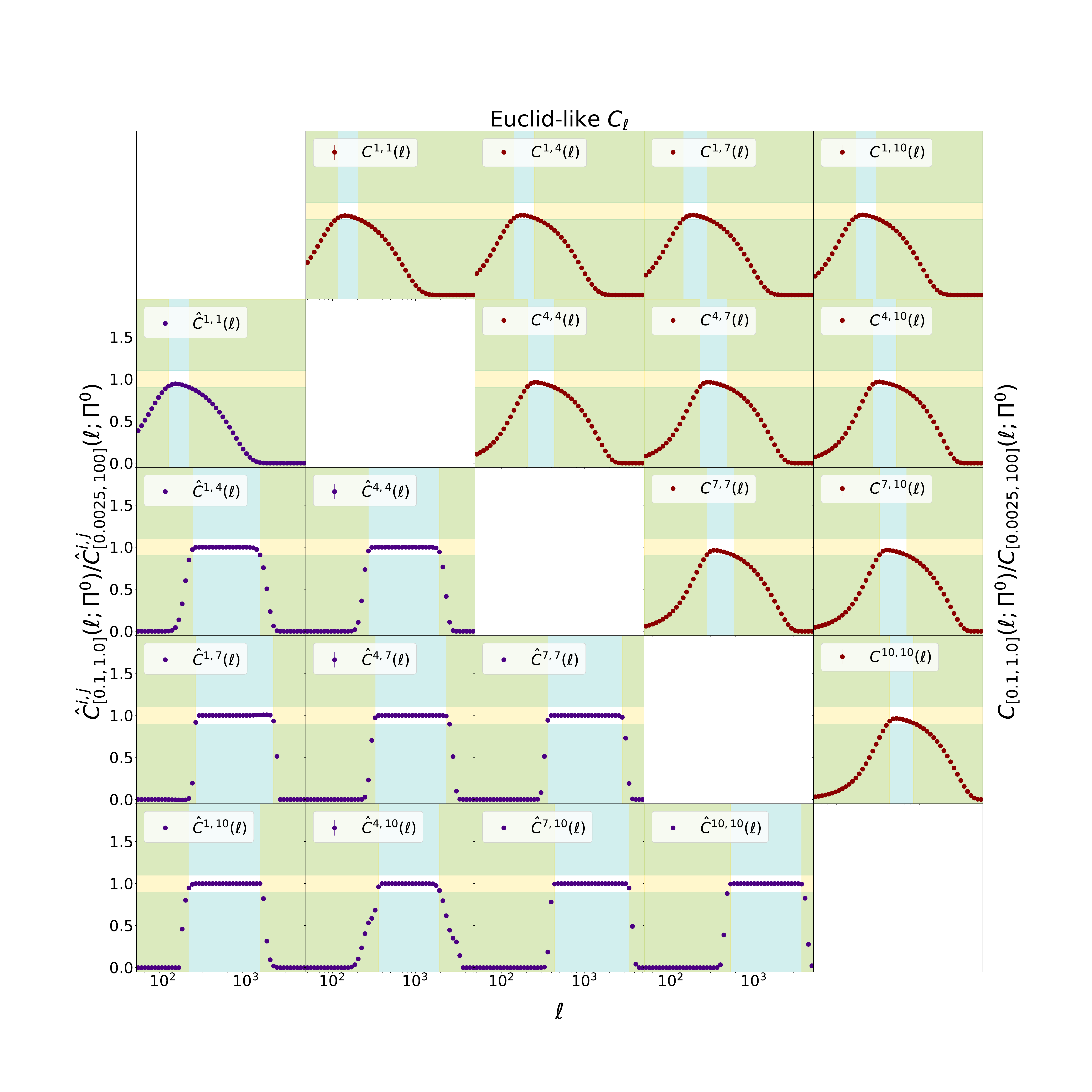}
\caption{Ratio of the data vector with $k_{\rm cut}$ to the full data vector (defined in Eq. \ref{Rfrak_def}) for the noBNT case $\fR^{(i,j)}(\ell;k_{\rm cuts},\Pi^0)$ (upper-right triangle) and the BNT case $\hfR^{(i,j)}(\ell;k_{\rm cuts},\Pi^0)$ (lower-left triangle). Each panel shows the ratio for the combination of tomographic bins indicated by $(i,j)$ (for the noBNT case) and $(a,b)$ (for the BNT case). For each tomographic bin, the curves are used to define which cut in $\ell$ corresponds to the constraint that modes $k>k_{\rm high}$ and $k<k_{\rm low}$ should not bias the data vector at a level exceeding $\mathcal{T}_\mathrm{FD}$ for $\ell_{\rm low}^{{(i,j)}}<\ell<\ell_{\rm high}^{{(i,j)}}$ for noBNT (resp. $\hat{\ell}_{\rm low}^{{(a,b)}} < \ell<\hat{\ell}_{\rm high}^{{(a,b)}}$ for BNT). All plots are given for $k_\mathrm{cuts} = (0.1\;{\rm Mpc^{-1}},1.0\;{\rm Mpc^{-1}})$, and the horizontal bright band indicates the fractional threshold $\mathcal{T}_\mathrm{FD} = \pm 0.1$. For each tomographic bin combination, the cut in $\ell$ is shown as the boundary of the cyan and green regions.
}
\label{fig:Cl}
\end{figure*}

We adopt the same approach as in paper I, where we determined which $\ell$-scale cut $\ell < \ell_{\rm cut}$ corresponds to a physical $k$-cut $k < k_{\rm cut}$ for a given accuracy. The difference here is that we now consider a domain cut, $k_\mathrm{cuts} = [k_\mathrm{low}, k_\mathrm{high}]$, and attempt to match $k_\mathrm{cuts}$ to a corresponding range in $\ell$-space such that the resulting data vector remains accurate to a specified precision within this $\ell$-space interval.
The relationship between $k$ and $\ell$ cuts for both of the BNT and noBNT data vectors will be estimated using these dimensionless ratios:

\bea
\hfR^{(a,b)} (\ell;k_{\rm cuts},\Pi^0) &\equiv& \frac{\hmC^{a,b}_{[k_\mathrm{low},k_\mathrm{high}]}(\ell;\Pi^0)}{\hmC^{a,b}_{[k_\mathrm{min},k_\mathrm{max}]}(\ell;\Pi^0)} \nonumber\\
{\fR}^{(i,j)}(\ell;k_{\rm cuts},\Pi^0)&\equiv&\frac{{\cal C}^{i,j}_{[k_\mathrm{low},k_\mathrm{high}]}(\ell;\Pi^0)}{{\cal C}^{i,j}_{[k_\mathrm{min},k_\mathrm{max}]}(\ell;\Pi^0)}
\label{Rfrak_def}
\eea

where $k_\mathrm{min} = 0.0025$ Mpc$^{-1}$ and $k_\mathrm{max} = 100$ Mpc$^{-1}$ is the entire computational domain in $k$-space. The $\hfR$ ratios represent the fractional difference between the angular power spectrum with $k$-cut (numerator) and no $k$-cut (denominator) for a given $\ell$. It literally captures the fractional power surviving the $k$-cut in $\ell$ space.
As noted in Paper I, the BNT transform does not need to be computed with the true cosmology. 
Actually, searching for the true cosmology for which nulling would be exact 
can itself be exploited (\cite{Touzeau2025BNT1,Touzeau2025BNT2}).

Following Paper I, we also define a fractional difference threshold $\tFD$ to characterise how a $\ell$-cut is defined based on the tolerance region with $\hfR = 1\pm\tFD$.  {The parameter $\tFD$ is set by the user, based on the bias level of the angular spectrum that can be tolerated (this value is the same for all elements of the data vector). This is in general determined by the accuracy required for a given cosmological parameter (e.g. $S_8$) to distinguish different cosmological scenario or independent measurements (e.g. Planck versus lensing measurements of $S_8$)}.
In order to include the impact of both $k_\mathrm{cuts}$ and $\tFD$, we use the same continuous Boolean weight function $\mathcal{V}(k_\mathrm{cuts},\tFD)$ as the paper I:

\be \label{eqn:ellcut}
{\mathcal{V}}(k_\mathrm{cuts},\mathcal{T}_\mathrm{FD}) = \begin{cases}
1, |\fR (k_{\rm cuts},\Pi^0)-1| \leq \mathcal{T}_\mathrm{FD} \\
0, |\fR (k_{\rm cuts},\Pi^0)-1| > \mathcal{T}_\mathrm{FD},
\end{cases}
\ee
where $\fR (k_{\rm cuts},\Pi^0)$ is the dimensionless power spectra ratio computed by the entire tomographic data vector, and can be applicable for the cases of noBNT and BNT. This Boolean weight function is used in the likelihood computation:
\bea
\Delta {\cal D}^* &\equiv& \mathcal{V}(k_\mathrm{cuts},\mathcal{T}_\mathrm{FD})\left({\cal C}(\Pi)-{\cal C}(\Pi^0)\right) \nonumber\\
\mathbb{C^*} &\equiv& \mathcal{V}(k_\mathrm{cuts},\mathcal{T}_\mathrm{FD}) \cdot \mathbb{C} \cdot \mathcal{V}(k_\mathrm{cuts},\mathcal{T}_\mathrm{FD})^{^T} \nonumber \\ 
\chi^2(\Pi;k_\mathrm{cuts},\mathcal{T}_\mathrm{FD})&=&\Delta {{\cal D}^*}^T \cdot (\mathbb{C^*})^{-1} \cdot \Delta {\cal D}^*\label{fb:chi2kFD}
\eea
where ${\cal C}(\Pi)$ is the entire tomographic data vector that is organized like:
\be \label{eqn:tomo}
{\cal C}_{[k_1,k_2]}(\Pi) = \begin{pmatrix}
    \begin{array}{cc}
        \begin{bmatrix}
        C^{1,1}_{[k_1,k_2]}(\ell;\Pi)  \\
       \end{bmatrix} {\rm all}~\ell's \\
       \begin{array}{c}
       \end{array}\\
       \begin{bmatrix}
        C^{1,2}_{[k_1,k_2]}(\ell;\Pi) \\
       \end{bmatrix} {\rm all}~\ell's \\
       \begin{array}{c}
       \end{array}\\
       \begin{bmatrix}
        C^{1,3}_{[k_1,k_2]}(\ell;\Pi) \\
       \end{bmatrix} {\rm all}~\ell's \\
        ...& \\
       \begin{array}{c}
       \end{array}\\
       \begin{bmatrix}
        C^{9,10}_{[k_1,k_2]}(\ell;\Pi) \\
       \end{bmatrix} {\rm all}~\ell's \\
       \begin{array}{c}
       \end{array}\\
       \begin{bmatrix}
        C^{10,10}_{[k_1,k_2]}(\ell;\Pi) \\
       \end{bmatrix} {\rm all}~\ell's \\
    \end{array}
\end{pmatrix}.
\ee
 
The $C^{i,j}_\ell$ or $\hat{C}^{a,b}_\ell$ difference is used for distinguish $i,j$ (noBNT) or $a,b$ (BNT) tomographic bins, and $\mathbb{C}$ is the corresponding covariance matrix.

Figure \ref{fig:Cl} shows the impact of a particular $k$-cut on the angular power spectrum $C_\ell$ for different tomographic combinations. It results in a biased $C_\ell$ for a range of $\ell$ which depends on the tomographic bin. As expected, the noBNT $C_\ell$ is significantly more biased than the BNT one. In order to quantify this bias, we define, for BNT, $\hat\ell^{(a,b)}_{\rm cuts}=(\hat\ell^{(a,b)}_{\rm low},\hat\ell^{{(a,b)}}_{\rm high})$
(and $\ell^{(i,j)}_{\rm cuts}$ for noBNT) as the range of $\ell$ where the bias does not exceed $\tFD$, i.e. where $\hfR^{(a,b)} (\ell \in \hat\ell^{a,b}_{\rm cuts};k_{\rm cuts},\Pi^0) \in (1 - \tFD,1+\tFD)$.

For each tomographic bin, the vertical blue regions on Figure \ref{fig:Cl} shows how $\hat\ell^{(a,b)}_{\rm cuts}$ (resp. $\ell^{(i,j)}_{\rm cuts}$ for noBNT) for $k_{\rm cuts}=(0.1,1.0)\;{\rm Mpc^{-1}}$ and $\mathcal{T}_\mathrm{FD} = 0.1$. The horizontal yellow bands shows the $1\pm \mathcal{T}_\mathrm{FD}$ region for $\hfR^{(a,b)}$ (resp. $\fR^{(i,j)}$). The blue area shows the range $\hat\ell^{{(a,b)}}_{\rm low} < \ell < \hat\ell^{{(a,b)}}_{\rm high}$ (resp. $\ell^{{(i,j)}}_{\rm low} < \ell < \ell^{{(i,j)}}_{\rm high}$) meets the $\mathcal{T}_\mathrm{FD}$ criteria and the green area shows the range of $\ell$ that must be removed in order to meet the criteria. This figure illustrates the impact of choosing a $k_\mathrm{cuts}$ as a $k$-band on the 2-D angular power spectrum and enables the definition of an $\ell$-band which approximates the effect of such a $k$-band. It is worth to note again here that the BNT ratios $\hfR^{(a,b)} (\ell;k_{\rm cuts},\Pi^0)$ at the transition between low and high $\ell$ is much sharper than the noBNT ratios ${\fR}^{(i,j)}(\ell;k_{\rm cut},\Pi^0)$ in both ends, which indicates the a $k$-space cut impacts many more $\ell$ modes in the noBNT data vector than the BNT one. Therefore, a definition of an arbitrary $\tFD$ is critical for noBNT analysis, but not so important for BNT-based analysis.

\subsection{Tomographic redshift bins cuts}
\label{subsec:IIIB}
In addition to applying $k$-band cuts, we can also perform the measurement within a subset of redshift bins. The motivation for doing these measurements in selected redshift bins is to verify the consistency of the cosmological parameter inference across bins. This is particularly relevant here because the redshift dependences of IA and lensing differ. Using the Euclid-like mock with ten redshift bins, we partition the lensing kernels into three groups: $\mathbb{Z}_{1,2,3}$ for the first three kernels $\{1,2,3\}$, $\mathbb{Z}_{4,5,6}$ for kernels $\{4,5,6\}$, and $\mathbb{Z}_{7,8,9,10}$ for kernels $\{7,8,9,10\}$. This partition is applied to both the non-BNT and BNT data vectors. For BNT, it is applied to the BNT-transformed lensing kernels $(a,b)$ in $C^{a,b}(\ell)$, whereas for the non-BNT case it is applied to the non-BNT lensing kernels $(i,j)$ in $C^{i,j}(\ell)$.

This binning strategy is designed to assess any possible inconsistency between redshift bins. For each lensing-kernel group, we include in the data vector all auto- and cross-correlations for which both indices e.g. $(a,b)$ for BNT, belong to the same group, and exclude any cross-correlations whose indices come from different groups. For example, under this scheme, the spectrum $C^{2,3}(\ell)$ is included in $\mathbb{Z}_{1,2,3}$, while $C^{3,4}(\ell)$ is excluded everywhere because its indices belong to different groups ($\mathbb{Z}_{1,2,3}$ and $\mathbb{Z}_{4,5,6}$).

\subsection{Analysis setup}
\label{subsec:IIIC}

\begin{table}[t]
\centering
\caption{\label{tbl:prior}%
The prior of all the parameters, except for IA parameters, used in our \textsc{Nautilus} sampling. $\mathcal{U}(a,b)$ is a flat prior 
between $a$ and $b$, and $\mathcal{N}(c,d)$ is a Gaussian prior that centred at $c$ with standard deviation $d$.}
\rowcolors{2}{white}{lightgray!20}
\begin{tabular}{|l|c|c|}
\hline
\rowcolor{cyan!10} 
\ \textrm{Parameters} \ &
\ \textrm{Fiducial Value} \ & 
\textrm{Priors} \\
\hline
\ $\Omega_m$ & $0.2905$ & $\mathcal{U}(0.1,0.6)$\\
\ $10^9 A_s$ & $2.1868$ & $\mathcal{U}(1.5,5.0)$\\
\ $\Omega_b$ & $0.0473$ & $\mathcal{U}(0.03,0.07)$\\
\ $n_s$ & $0.9690$ & $\mathcal{U}(0.92,1.02)$\\
\ $h$ & $0.6898$ & $\mathcal{U}(0.55,0.85)$\\
\ $A_\mathrm{B}$ & $3.1300$ & $\mathcal{U}(1.0,6.0)$\\
\hline
\hline
\ $\sigma_8$ & $0.8256$ & \ Derived Parameter \ \\
\ $S_8$ & $0.8124$ & \ Derived Parameter \ \\
\hline
\end{tabular}
\end{table}

The analysis presented here extends our Paper I study by incorporating intrinsic alignment and applying a narrower $k$-cut, while retaining the same overall objective: to compare the BNT and non-BNT cases in terms of (1) the precision of the cosmological parameter measurements, and (2) the bias of the posterior relative to the input model (including cases where the intrinsic alignment model is invalid). We carry out a series of Bayesian posterior estimations to assess the impact of intrinsic alignment, assuming a "True Background Cosmology" (TBC) and employing two distinct $P(k)$ models: the canonical \textsc{HMcode} \cite{HMcode16} and the non-standard \textsc{AxionHMcode} \cite{AxionHMcode}, which accounts for the effects of ultralight axions on the small-scale growth of large-scale structure. In this study, we adopt a fixed axion mass of $m_\mathrm{axion} = 10^{-24}\,$eV and an axion density of $\Omega_\mathrm{axion} = 0.15 \times \Omega_m = 0.0436$.

The detailed cosmological parameters for our TBC are listed in Table~\ref{tbl:prior}. The fiducial values and priors of these parameters remain identical across all posterior samplings. For the intrinsic alignment component, we consider three different models. The NLA and TATT model parameters follow the fits reported in \cite{Samuroff2021HydroIA}, derived from galaxy correlation functions based on the IllustrisTNG hydrodynamical simulations \cite{Springel2018TNG, Nelson2019TNG}. In addition, the NLA-z model adopts the best-fit parameters from the KiDS Legacy cosmic shear analysis \cite{Wright2025KiDSLegacy}.

\begin{table}[t]
\centering
\caption{\label{tbl:prior_IA}%
The prior of all the parameters used in our \textsc{Nautilus} sampling. For table width considerations, we use a simplified notation here $\mathcal{U}_{a,b} = \mathcal{U}(a,b)$ for a flat prior 
between $a$ and $b$. The single value prior means we fix that parameter to this particular value in that prior. We don't necessarily sample all parameters below in every chain. If they are not sampled, their values are set to the 'Fiducial Value' column which indicates the values taken by the parameters for the fiducial cosmology.
}
\rowcolors{2}{white}{lightgray!20}
\begin{tabular}{|l|c|c|c|c|c|c|}
\hline
\rowcolor{cyan!10} 
\ \textrm{Parameters} \ &
\ $A_\mathrm{TA}$ \ & 
\ $\eta_\mathrm{TA}$ \ & 
\ $B_\mathrm{TA}$ \ & 
\ $b_\mathrm{TA}$ \ & 
\ $A_\mathrm{TT}$ \ & 
\ $\eta_\mathrm{TT}$ \\
\hline
\ \textrm{NLA Model} \ &
\ $1.71$ & $0.0$ & $-$ & $-$ & $-$ & $-$ \\
\ \textrm{TATT Model} \ &
\ $1.29$ & $0.0$ & $-$ & $0.21$ & $0.32$ & $0.0$ \\
\ \textrm{NLA-z Model} \ &
\ $0.3$ & $-$ & $-3.7$ & $-$ & $-$ & $-$ \\
\hline
\ \textrm{NLA Prior} \ &
\ $\mathcal{U}_{-6,6}$ & $\mathcal{U}_{-5,5}$ & $-$ & $-$ & $-$ & $-$ \\
\ \textrm{TATT Prior} \ &
\ $\mathcal{U}_{-6,6}$ & $\mathcal{U}_{-5,5}$ & $-$ & $0.0$ & $\mathcal{U}_{-6,6}$ & $\mathcal{U}_{-5,5}$ \\
\ \textrm{NLA-z Prior} \ &
\ $\mathcal{U}_{-6,6}$ & $-$ & $\mathcal{U}_{-6,6}$ & $-$ & $-$ & $-$ \\
\hline
\end{tabular}
\end{table}
 
In the first part of our investigation, we assess the stability of the BNT transform under the assumption that we \textit{know the IA model precisely.} For each of the NLA, TATT, and NLA-z models, we perform posterior sampling using the corresponding prior—i.e., the NLA Prior is run against the fiducial data vector generated with the NLA Model, and likewise for the others. As in Paper I, nonlinear modeling is handled by performing the sampling with \textsc{HMcode}, while the fiducial data vector is generated with \textsc{AxionHMcode}.

In the second part of our analysis, we examine whether the BNT transform can shed light on the validity of assuming that the IA model is accurately known. To this end, we perform sampling under the NLA model while calibrating against fiducial data vectors generated with models that include additional degrees of freedom—specifically, the NLA-z and TATT models. To reduce complexity, both the sampled and fiducial data vectors in this part are generated using the same nonlinear prescription: \textsc{HMcode}. This setup allows us to test the IA model independently of whether the underlying cosmology is assumed to be correct, a possibility enabled by the flexibility of the BNT transform, which permits new and informative manipulations of the data vector.

In addition to the above, the following settings are used for all MCMC runs:

\begin{itemize}
    \item We use 50 logarithmically spaced $\ell$-bins, with bin centers ranging from $\ell_\mathrm{min} = 50$ to $\ell_\mathrm{max} = 5000$, and adopt the same redshift distribution as in paper I, which also yields the kernels shown in Figure~\ref{fig:nz}.
    
    \item All data vectors used in the posterior sampling are computed with the Core Cosmology Library (CCL) \cite{Pyccl}, a publicly available and standardized tool for calculating cosmological observables. CCL is developed and maintained by the LSST Dark Energy Science Collaboration (DESC) \cite{DESC2018DESC}.
    
    \item The covariance matrix is computed using \textsc{OneCovariance} \cite{OneCov}, a tool optimized for estimating covariance matrices from two-point statistics in photometric large-scale structure surveys. In our calculation, the statistical noise components are derived assuming an effective survey area of $A_\mathrm{eff} = 15000\,\mathrm{deg}^2$, a galaxy number density per tomographic bin of $n_{\rm gal} = 3.0,\mathrm{arcmin}^{-2}$, and a shape noise of $\sigma_e = 0.3$ \cite{Euclid2020Forecast}.
    
    \item Parameter posteriors are obtained with the \textsc{Nautilus} package \cite{Nautilus}, a highly efficient importance nested-sampling toolkit for Bayesian posterior and evidence estimation.
\end{itemize}

\section{Results}
\label{sec:results}

\subsection{BNT and Intrinsic Alignment}

We begin by analyzing cases where the intrinsic-alignment (IA) model used to generate the fiducial data vector is also employed in the sampling model, but with different parameter values in each. In the sampling, each parameter is either fixed or varied under the “TATT Prior,” using fiducial data vectors generated with the “TATT Model.” (see Table \ref{tbl:prior_IA}). Note that the $b_\mathrm{TA}$ parameter is fixed to different values in these two setups, so the “TATT Prior” does not encompass the “TATT Model.” This is because $b_\mathrm{TA}$ is strongly degenerate with $A_\mathrm{TT}$ (see Eq. \ref{eq:TATTparams}). This configuration therefore represents a case where the model is well constrained but not perfectly known—an ideal test of the BNT transform’s stability in the presence of intrinsic-alignment signals.
Throughout this section, fiducial data vectors are computed with \textsc{AxionHMcode}, whereas the sampling itself is performed with the standard, no-axion version of \textsc{HMcode}.

\begin{figure*}
\centering
\includegraphics[width=0.85\textwidth, trim = 0.25cm 0.2cm 0.2cm 0.25cm, clip]{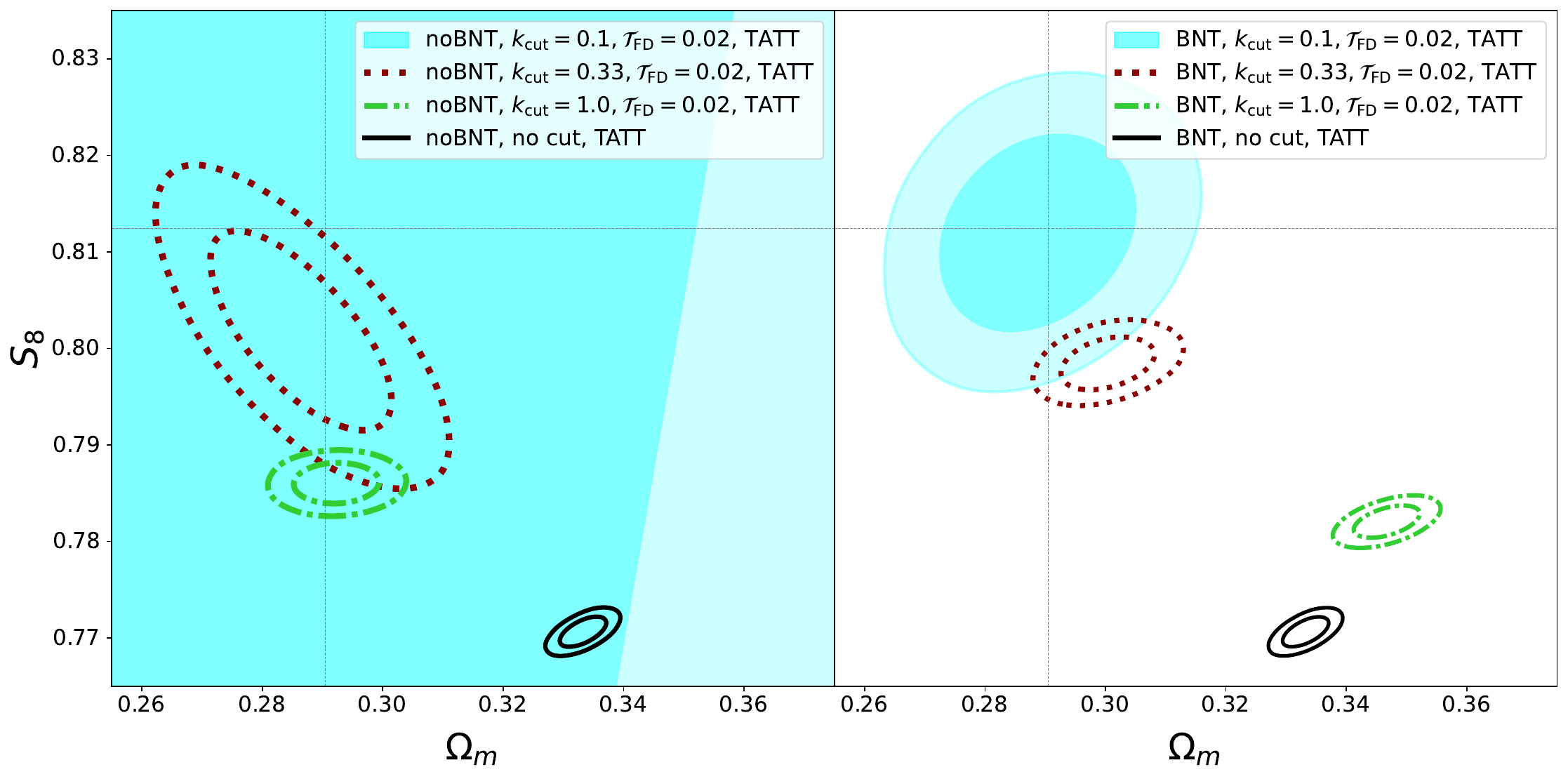}
\caption{Constraints in the $(\Omega_m,S_8)$ plane for the noBNT (left panel) and BNT (right panel). The thin solid lines indicate the values of the fiducial cosmology. The power spectrum of all models during the sampling process is calculated using the \citet{HMcode16}-version of \textsc{HMcode}, while the fiducial is calculated using the \textsc{AxionHMcode}\cite{AxionHMcode}. The fiducial also included a \citet{Samuroff2021HydroIA}-fit of TATT model from the IllustrisTNG galaxies. Four posterior contours are represented: the {\color[HTML]{000000}\bf Black} solid line corresponds to the case where no $k_{\rm cut}$ and no $\mathcal{T}_\mathrm{FD}$ are applied. The {\color[HTML]{31CD31}\bf Lime Green}, {\color[HTML]{8B0000}\bf Dark Red} and {\color[HTML]{00AFFF}\bf Cyan} contours correspond to $k_{\rm cut}=1.0, 0.33, \mathrm{and\ } 0.1\;{\rm Mpc^{-1}}$ respectively, and fixed $\mathcal{T}_\mathrm{FD}=0.02$. Note that the {\color[HTML]{00AFFF}\bf Cyan} contours for noBNT and $k_{\rm cut}=0.1\;{\rm Mpc^{-1}}$ exceeds the plot boundary.}
\label{fig:kcut_TATT_OmS8}
\end{figure*}
 
We first examine the joint constraints on $(S_8, \Omega_m)$ in a setting where the fiducial nonlinear power spectrum is biased relative to the model used for sampling. As in Paper I, we compare constraints from the BNT and noBNT approaches for various choices of $k_{\rm cut}$, using \textsc{AxionHMcode} with non-zero intrinsic alignments for the fiducial data vector, while the sampling model uses \textsc{HMcode}. In Figure~\ref{fig:kcut_TATT_OmS8}, we show the cosmological constraints for different physical scale cuts $k_{\rm cut}$, sampled over a TATT model with $b_\mathrm{TA} = 0.0$, compared to the fiducial TATT model with $b_\mathrm{TA} = 0.21$.
Notably, this mismatch in $b_\mathrm{TA}$ does not introduce any significant additional bias in the cosmological constraints. The bias in $S_8$ and $\Omega_m$ appears when small-scale information is included, as shown by the shift of contours in Figure~\ref{fig:kcut_TATT_OmS8}, which is primarily due to the impact of ultralight axions. However, as we progressively restrict the analysis to larger scales, the $S_8$ and $\Omega_m$ constraints converge toward the “true background cosmology” indicated by the cross in the contour plot. At $k_\mathrm{cut} = 0.1\;\mathrm{Mpc}^{-1}$, the recovered cosmology aligns well with the fiducial. These results closely match the result in Paper I, where intrinsic alignment was not included, meaning that the biases seen in Figure~\ref{fig:kcut_TATT_OmS8} are not caused by the biased IA model, but effectively by the incorrect non-linear model of the mass power spectrum. We conclude that the performance of BNT is not degraded by the presence of intrinsic alignment. By contrast, the noBNT constraints lose most of their constraining power in this regime, a result also similar to Paper I. These results suggest that, even with good but imperfect knowledge of the IA model, the BNT-transform-based $\ell$-cut method remains effective for delivering unbiased cosmological constraints.

\begin{figure*}
\centering
\includegraphics[width=0.85\textwidth, trim = 0.25cm 0.2cm 0.2cm 0.25cm, clip]{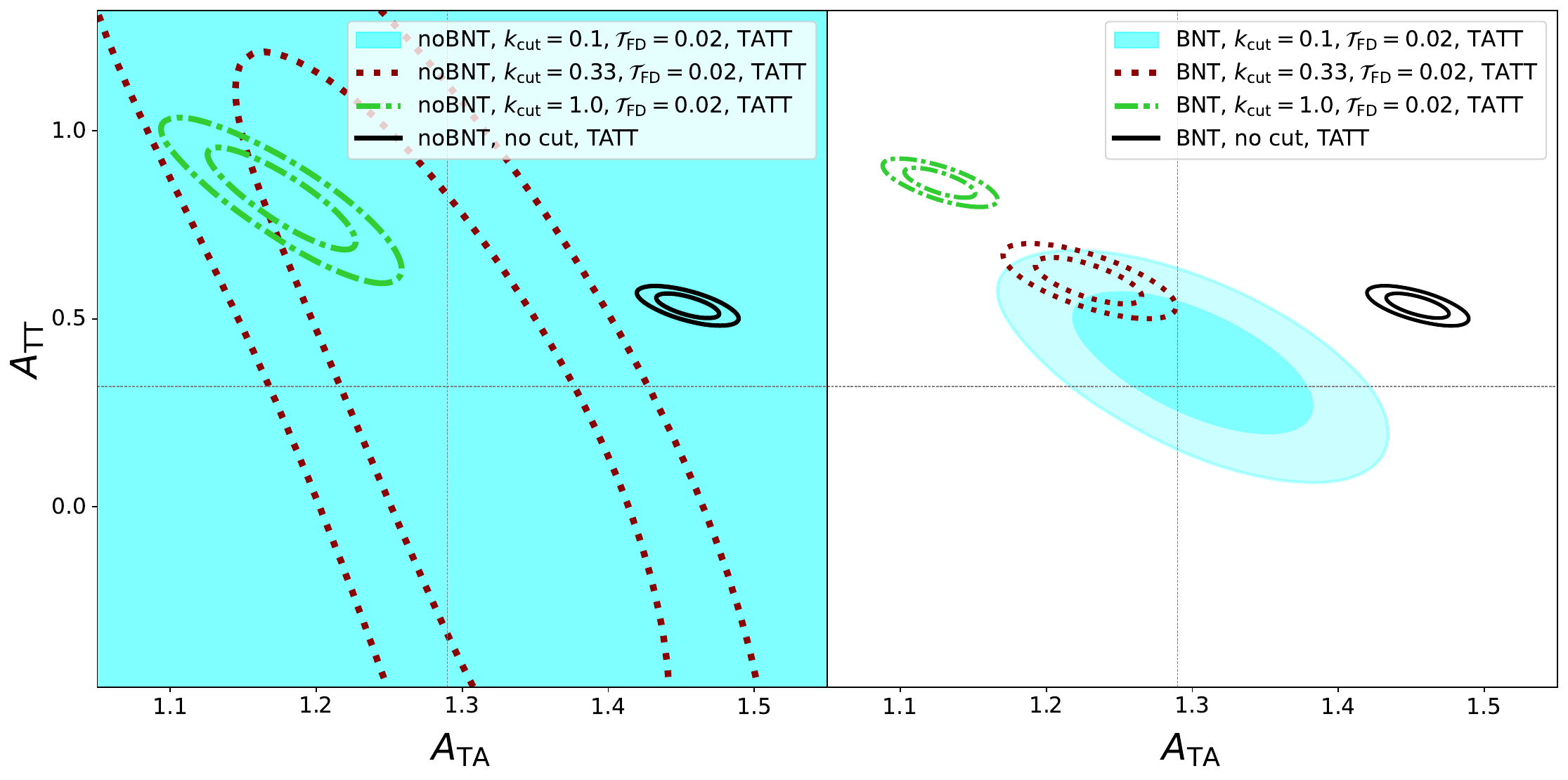}
\caption{Same to the Figure \ref{fig:kcut_TATT_OmS8}, but showing the constraints of $A_\mathrm{TA}$ and $A_\mathrm{TT}$ instead of $\Omega_m$ and $S_8$.}
\label{fig:kcut_TATT_TATT}
\end{figure*}

Having a different $b_\mathrm{TA}$ for the fiducial and sampling models has a significant impact on the IA modeling. This is illustrated in Figure~\ref{fig:kcut_TATT_TATT} which shows the joint constraints on $A_\mathrm{TA}$ and $A_\mathrm{TT}$ using the same posterior as in Figure~\ref{fig:kcut_TATT_OmS8}. As expected, the noBNT posteriors exhibit a tight and biased constraint with large $k_\mathrm{cut}$ values, but give weak constraints when using smaller $k_\mathrm{cut}$ values. By contrast, the BNT posterior not only retain its constraining power on cosmological parameters but it also retains its constraining power on $A_{\rm TT}$ and $A_{\rm TA}$ despite a biased $b_\mathrm{TA}$ value in the sampling models.
Although the location of the maximum posterior in the $(A_\mathrm{TA}, A_\mathrm{TT})$ plane does not move monotonically with changing the scale cut, we recover a well-constrained $A_\mathrm{TT}$ value at large-scale $k_\mathrm{cut}$ using the BNT transform.

\subsection{Biases induced by Intrinsic Alignment Models}
\label{subsec:IVB}

It is possible that the IA model adopted in the sampling is not fully representative of reality. The key question is therefore: to what extent can we trust the cosmological parameter inference, and what diagnostic tools are available to assess the validity of the assumed IA model? An incorrect IA model can, in general, introduce biases either in redshift or in $k$-space. The BNT transform provides access to $k$-space, while redshift-space information can be obtained through $z$-bin grouping, as discussed in Section~\ref{subsec:IIIB}. In the following, we compare the non-BNT and BNT cases for different $k$-cuts and lensing-kernel groupings.

\begin{figure*}
\centering
\includegraphics[width=0.85\textwidth, trim = 0.25cm 0.2cm 0.2cm 0.25cm, clip]{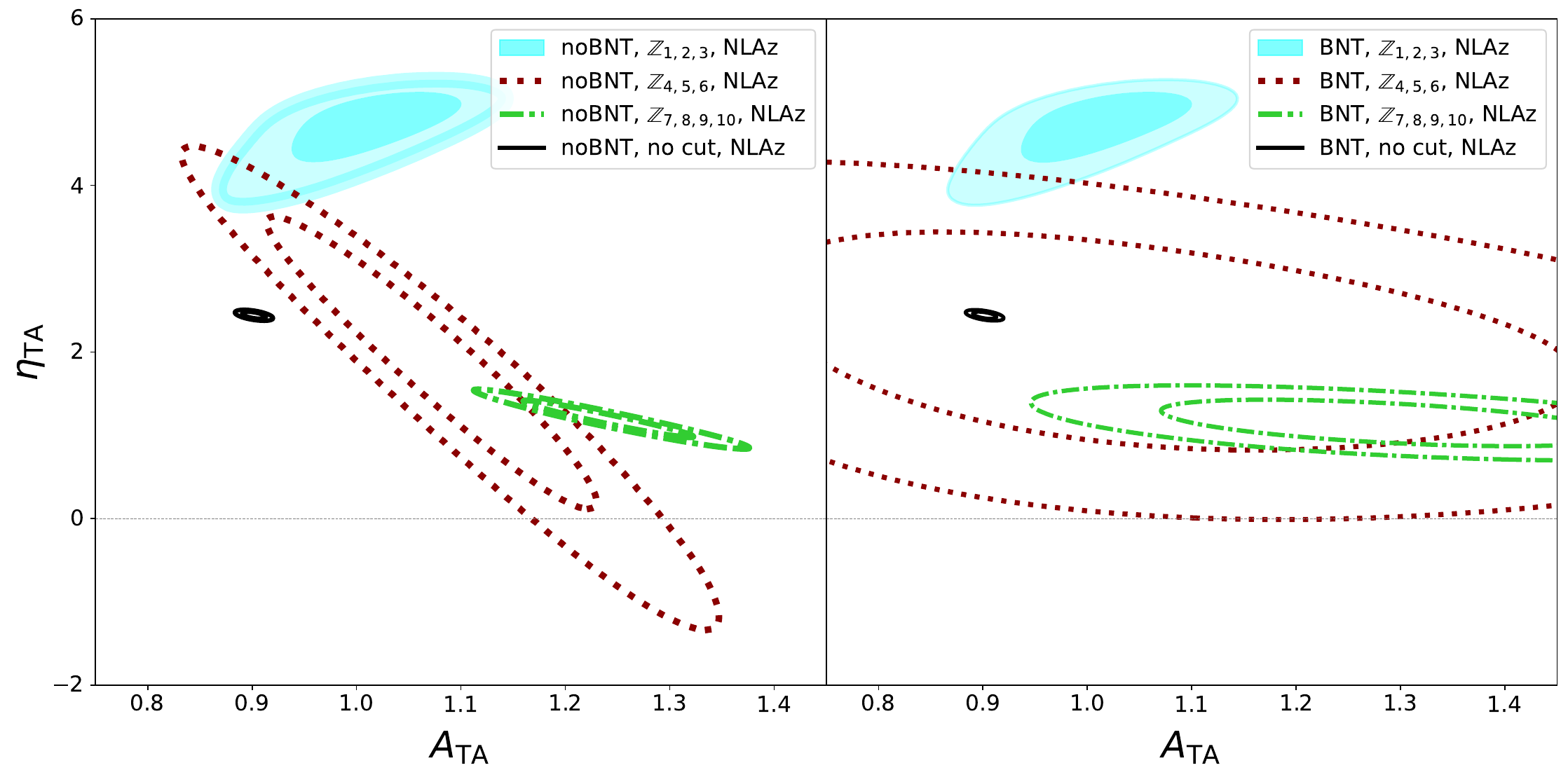}
\caption{Constraints in the $(A_\mathrm{TA},\eta_\mathrm{TA})$ plane for the noBNT (left panel) and BNT (right panel). Note that the fiducial value of $A_{\rm TA}$ is $1.71$ (see Table \ref{tbl:prior_IA}) and therefore outside the x-axis range. The power spectrum of all models for sampling and fiducial are both calculated using the \citet{HMcode16}-version of \textsc{HMcode}. The fiducial also included a \citet{Wright2025KiDSLegacy}-fit of NLA-z model from the KiDS-Legacy data, but we fit with an NLA Model against it in our sampling. Four posterior contours are represented: the {\color[HTML]{000000}\bf Black} solid line corresponds to the case where no $z_{\rm cuts}$ is applied. The {\color[HTML]{00AFFF}\bf Cyan}, {\color[HTML]{8B0000}\bf Dark Red}, and {\color[HTML]{31CD31}\bf Lime Green} contours correspond to the case where we only used the data vector from first three tomographic bins, $4^{\rm th}-6^{\rm th}$ tomographic bins, and last four tomographic bins.}
\label{fig:zBand_NLAz_TAS8}
\end{figure*}

We first investigate the impact of an incorrect IA model in the redshift space by using the NLA-z model from Table \ref{tbl:prior_IA} for fiducial and NLA model for the sampling. The NLA-z model introduces a different redshift dependence than the standard NLA model, which illustrates the impact of choosing an incorrect IA model. The data vector is split in three tomographic redshift groups: $(i,j)$ or $(a,b)$ for noBNT or BNT, belonging to either $\{1,2,3\}$, $\{4,5,6\}$, or $\{7,8,9,10\}$. For each subset, we exclude cross-correlations between bins belonging to different groups, as explained in Section \ref{subsec:IIIB}. Figure~\ref{fig:zBand_NLAz_TAS8} shows the resulting constraints on $A_\mathrm{TA}$ and $\eta_\mathrm{TA}$ derived from the posterior sampling.
In both the noBNT and BNT analyses, the resulting contours display substantial inconsistencies across different redshift-bin groupings. Constraints from the first three tomographic bins deviate from those obtained using the full redshift range by more than $5\sigma$ in $\eta_\mathrm{TA}$ ($7.01\sigma$ for BNT and $7.03\sigma$ for noBNT using the mahalanobis distance \cite{mahalanobis36distance} between the posterior means on the multi-dimensional parameter space). Likewise, the constraints from the last four tomographic bins differ from those of the first three by over $10\sigma$ (10.85 for BNT and 12.19 for noBNT), while still showing a tension exceeding $5\sigma$ (7.16 for BNT and 8.36 for noBNT) relative to the unsliced case. These findings demonstrate that strong internal inconsistencies can arise when the IA model is incorrect. Redshift grouping therefore provides a useful diagnostic for testing the validity of the assumed IA model.

\begin{figure}[htbp]
\centering
\includegraphics[width=0.49\textwidth, trim = 0.1cm 0.0cm 0.0cm 0.0cm, clip]{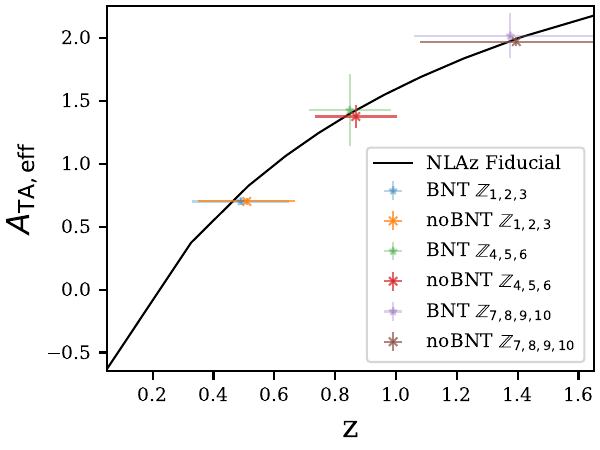}
\caption{The effective tidal alignment amplitude $A_\mathrm{TA,eff}(z)$ (Eq. \ref{eqn:A_ta_eff}) constraints of all three redshift slices derived from the contour of the Figure~\ref{fig:zBand_NLAz_TAS8}, for both BNT and noBNT cases. The z-values and their errorbars of the data points are derived from the weighted mean and weighted standard deviation, when all galaxies within the $z$-slice (like $\mathbb{Z}_{1,2,3}$) are categorized to be within one single redshift distribution. For the illustration purpose only, the z-value of the noBNT and BNT data points are each shifted by $\pm 0.01$ on the plot, with their $A_\mathrm{TA,eff}(z)$ kept to its original constraint value.}
\label{fig:zBand_NLAz_AIA}
\end{figure}

Based on the BNT and noBNT constraints shown in Figure~\ref{fig:zBand_NLAz_TAS8}, we fit an effective tidal alignment amplitude, $A_\mathrm{TA,eff}(z)$ (using the definition from Eq.\ref{eqn:A_ta_eff}) for each redshift grouping. Figure~\ref{fig:zBand_NLAz_AIA}  presents the resulting $A_\mathrm{TA,eff}(z)$ compared to the input NLA-z Fiducial model.
The centre values and bin width along the $z$-axis are obtained from the $n(z)$-weighted mean and standard deviation, by combining the redshift slices $\mathbb{Z}_{1,2,3}$, $\mathbb{Z}_{4,5,6}$, and $\mathbb{Z}_{7,8,9,10}$ into three effective single redshift distributions. This demonstrates that, for an intrinsic alignment model with explicit redshift dependence, such an approach can serve as a viable method to constrain the detailed redshift evolution of the IA model if the sampling model cannot recover it. For instance, $A_\mathrm{IA,eff}(z)$ (Eq. \ref{eqn:A_ta_eff}) cannot cross zero with our sampling IA model (Eq. \ref{eqn:c1}), but it can do so with the fiducial IA model (Eq. \ref{eq:nlaz}), therefore creating a discrepancy.

\begin{figure*}
\centering
\includegraphics[width=0.85\textwidth, trim = 0.25cm 0.2cm 0.2cm 0.25cm, clip]{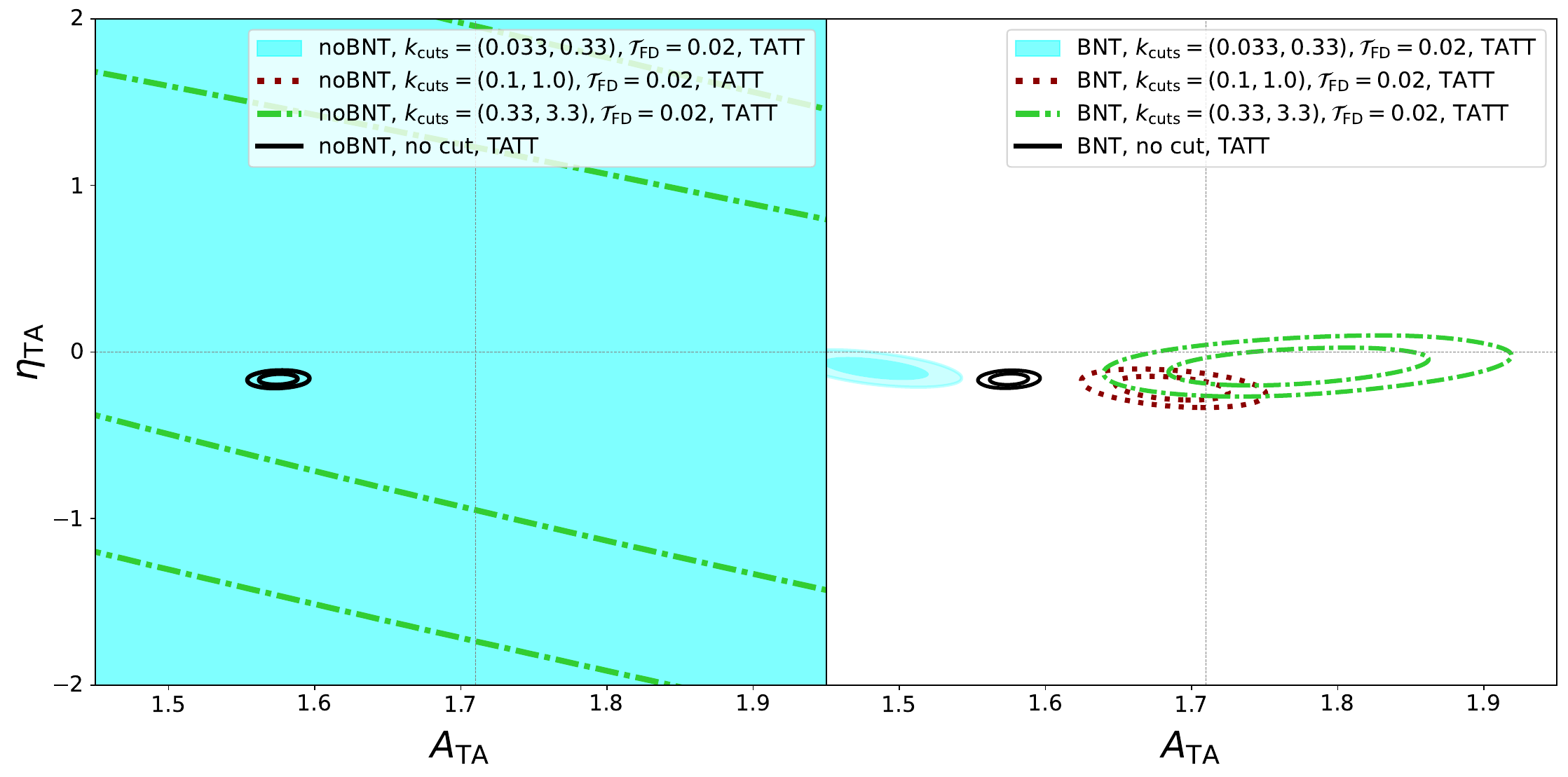}
\caption{Constraints in the $(A_\mathrm{TA},\eta_\mathrm{TA})$ plane for the noBNT (left panel) and BNT (right panel). The power spectrum of all models for sampling and fiducial are both calculated using the \citet{HMcode16}-version of \textsc{HMcode}. The fiducial also included a \citet{Samuroff2021HydroIA}-fit of TATT model from the IllustrisTNG galaxies, but we fit with an NLA Model against it in our sampling. Four posterior contours are represented: the {\color[HTML]{000000}\bf Black} solid line corresponds to the case where no $k_{\rm cuts}$ and no $\mathcal{T}_\mathrm{FD}$ are applied. The {\color[HTML]{31CD31}\bf Lime Green}, {\color[HTML]{8B0000}\bf Dark Red} and {\color[HTML]{00AFFF}\bf Cyan} contours correspond to $k_{\rm cuts}=(0.033,0.33)\;{\rm Mpc^{-1}}$, $(0.1,1.0)\;{\rm Mpc^{-1}}$, and $(0.33,3.3)\;{\rm Mpc^{-1}}$ respectively, and fixed $\mathcal{T}_\mathrm{FD}=0.02$. Note that all noBNT contours with $k_\mathrm{cuts}$ exceeds the plot boundary. The vertical highlight line on the $A_\mathrm{TA}$ gives the NLA fit in the same sample that derives the TATT model we use in the fiducial here.}
\label{fig:kBand_TATT_TAS8}
\end{figure*}

Similarly, we investigate the impact of $k$-band cuts when the fiducial and sampling IA models differ. In this test, we adopt the $k$-dependent nonlinear TATT model, which introduces additional clustering power at large $k$-scales compared to the conventional NLA model. The fiducial data vector is generated using the TATT model with parameters fitted from the \citet{Samuroff2021HydroIA} study, while sampling is performed with the standard NLA model. Figure~\ref{fig:kBand_TATT_TAS8} shows the constraints on ($\eta_{\rm TA},A_{\rm TA}$) for three $k$-bands: $k_\mathrm{cuts} = (0.033, 0.33)$, $(0.1, 1.0)$, and $(0.33, 3.3)\;\mathrm{Mpc}^{-1}$ . Notably, the noBNT analysis shows a substantial loss of constraining power, particularly for the intrinsic-alignment parameter $A_\mathrm{TA}$. In contrast, the contours of the BNT analysis are not increasing enormously, showing that the constraining power is preserved across all bands. The constraints on $S_8$ remain also consistent across all $k$ bins (not shown). On the other hand, the best fit posterior of $A_\mathrm{TA}$ strongly depends on the $k$ bin. There is an increasing systematic shifts from the fiducial value of the TATT model ($A_\mathrm{TA} = 1.32$) for the low $k$ bin, towards that of the NLA model fit to the same simulated galaxy sample ($A_\mathrm{TA} = 1.71$). This tension in $A_\mathrm{TA}$ between the $(0.033, 0.33)$ and $(0.33, 3.3)$ bands is $5.13\sigma$. This result demonstrates a clear scale-dependent inconsistency: the IA model used in the sampling pipeline fails to capture the scale dependence intentionally introduced in the fiducial -- exactly the discrepancy this test is designed to expose. This behaviour is not shared by the noBNT analysis.

\begin{figure}[htbp]
\centering
\includegraphics[width=0.49\textwidth, trim = 0.1cm 0.1cm 0.0cm 0.0cm, clip]{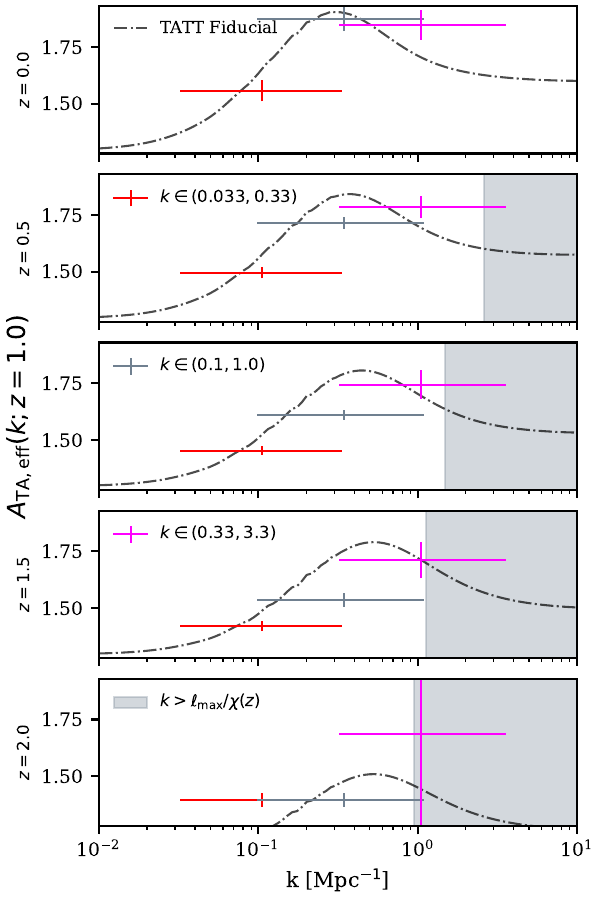}
\caption{The effective tidal alignment amplitude $A_\mathrm{TA,eff}(k,z)$ (Eq. \ref{eqn:A_ta_eff}) constraints of all three physical scale slices derived from the contour of the Figure~\ref{fig:kBand_TATT_TAS8}, for the constraints derived from BNT-transformed data vector at five different redshifts: $z = 0.0,\; 0.5,\; 1.0,\; 1.5,\; 2.0$. The errorbar on $k$-axis indicates the width of the $k$-band used in that data point, and the shaded region indicates the absolute maximum of $k$ we can explore due to our maximum $\ell_\mathrm{max} = 5000$ choice.}
\label{fig:kBand_TATT_AIA}
\end{figure}

As in Figure~\ref{fig:zBand_NLAz_AIA} for the redshift-dependent NLA model, we fit an effective amplitude $A_\mathrm{TA,eff}(k)$ within each of the $k$ bands.
However, because the TATT model predicts an $A_\mathrm{TA,eff}$ that varies with redshift, it remains necessary to examine its behavior across redshift when studying the $k$-dependence. Figure \ref{fig:kBand_TATT_AIA} shows the $A_\mathrm{TA,eff}(k)$ constraints obtained from a simple NLA model in three physical-scale slices, together with the dash-dotted curve representing the TATT model used in this analysis. To better illustrate the fits, we display$A_\mathrm{TA,eff}(k)$ at five representative redshifts: $z = 0.0,\; 0.5,\; 1.0,\; 1.5,\; 2.0$. Although the limited flexibility of our current model prevents it from fully reproducing the detailed $A_\mathrm{TA,eff}(k)$ behavior predicted by TATT, the results clearly show that the IA signal in the observational data vector exhibits a distinct $k$-dependence absent from the standard NLA model.

\section{Discussion and Conclusions}

In Paper I \cite{Gu2025BNT1}, we used the BNT transform to quantify which $\ell$-space truncations in each redshift-bin combination lead to a separation of physical scales in $k$-space. This truncation effectively selects the scale information retained in the cosmological inference; however, Paper I assumed the absence of intrinsic alignments.

Building on the paper I analysis, we examine the behaviour of the BNT transform in two distinct scenarios: one in which the assumed IA model is reasonably accurate but not exact, and another in which it fails to capture the true IA behaviour. We then assess whether the scale separation in $k$-space provided by the BNT method remains valid under these two scenarios.

In the first scenario, the “reasonably accurate” IA case corresponds to using the same IA model (TATT) for both the sampling and fiducial analyses, while certain fixed parameters of the fiducial model (TATT model in Table \ref{tbl:prior_IA}) cannot be recovered by the sampling model (TATT prior in Table \ref{tbl:prior_IA}). We find that the BNT approach still succeeds in separating the physical scales in $k$-space and yields an unbiased estimate of $S_8$. Quantitatively, with a $k$-cut of $k\le 0.1\,\rm Mpc^{-1}$ and a fractional threshold $\tFD=0.02$, BNT is able to recover an unbiased $S_8$ to better than 2\% while non-BNT looses all constraining power. With less stringent $k$-cuts ($k\le 0.33\,\rm Mpc^{-1}, k\le 1\,\rm Mpc^{-1}$), the $S_8$ bias can be significant, largelly exceeding 10$\sigma$ for the no $k$-cut version, for both BNT and non-BNT.

In the second scenario, the fiducial and sampling IA models differ fundamentally—either in redshift or in $k$-space—and can therefore introduce significant biases in the cosmological constraints. The key question is whether we can diagnose inconsistencies in the IA model and thus identify potential biases in the inferred cosmological parameters. We tested two scenarios: one where the fiducial model is NLA-z and the sampling model is NLA and the other where the fiducial is TATT and the sampling is NLA (see Table \ref{tbl:prior_IA} for the details).
Our results show that when the two IA models differ in redshift space, both the noBNT and BNT analyses provide an internal self-consistency check that reveals the discrepancy between the fiducial and sampling models. However, when the IA models differ in $k$-space, only the BNT approach provides the correct diagnostic, the noBNT approach does not provide any means of identifying an incorrect IA model. The discrepancies in redshift and $k$ bins are significant and exceed 5$\sigma$ ($5.13\sigma$ for $k$ bins and $>7 \sigma$ for $z$ bins). 

The main conclusion of this work is that intrinsic alignments do not cause the BNT method to fail. The separation of $k$-scales based on linear combinations of tomographic bins still effectively removes high-$k$ modes where the modelling is unreliable. We also explored the use of $k$-bins as an extension of BNT to isolate physical scales around specific features of interest. Our results show that a reasonably accurate IA model is required but need not be exact, and that when the IA model is fundamentally incorrect, BNT nonetheless provides a consistency test for identifying such cases. The robustness against biased IA models can be understood from Figures \ref{fig:zBand_NLAz_AIA} and \ref{fig:kBand_TATT_AIA}: in the BNT cases, the binning in $k$ or $z$ space allows the bias introduced by an IA model to be locally compensated within a given $k$ or $z$ bin by adjusting $A_{\rm TA,eff}$, without compromising the cosmological diagnostic power of the BNT approach.

{\bf
A natural extension of this work would be to investigate whether the IA model can be characterized non-parametrically using the discrepancies observed across the redshift and wavenumber bins. One could envision an iterative procedure that progressively corrects these discrepancies in both $z$ and $k$ space. We leave this for future study, but our results suggest that an iterative framework, combining cosmological-parameter inference with $k$-cut analyses and enforcing self-consistency across redshift and $k$ bins using the BNT transform, could help further constrain the IA model while simultaneously reducing non-linear modelling uncertainties.
}

These findings highlight the importance of employing the BNT methodology in Stage-IV lensing analyses to test and mitigate nonlinear and intrinsic-shape systematics, thereby enabling unbiased cosmological constraints with optimal control over scale cuts. This paper serves both as a proof of concept and as an illustration of a general consistency-testing framework. Follow-up studies are anticipated in the following directions:

\begin{itemize}
    \item The precision and the characterization of residual systematics of current Stage III cosmic shear surveys is improving, and the first data releases of Stage IV data are also not too far away. The KiDS-Legacy survey \cite{Wright2025KiDSLegacy}, with 6 tomographic bins, is a perfect Stage III survey benchmark for testing the BNT transform (Gu et al, {\it in preparation}).
    
    \item In paper I, we also proposed a mathematical framework to distribute the constraining power on different parameters on each band of the physical scale. This method still has great potential, like applying consistency checks on the survey setup, or being another method to perform a reconstruction of $P(k)$ with the BNT transform.
    
\end{itemize}

\begin{acknowledgments}
All theoretical calculations are based on \textsc{Pyccl}\cite{Pyccl} and we present our whole sampling code suit in \cite{BNT_repo}. 
We calculate errorbars and plot posterior contours plotting with \textsc{Getdist}\cite{getdist} package. We are grateful to Hendrik Hildebrandt, Mike Hudson, Bhuvnesh Jain, Benjamin Joachimi, Cora Uhlemann, Angus Wright, Ziang Yan, and the whole German Centre for Cosmological Lensing (GCCL) for their constructive feedback during the construction of the entire structure for the study and the analysis. We thank Sophie Vogt for her help with \textsc{AxionHMcode}. We thank Robert Reischke for providing his covariance matrix calculation code \textsc{OneCovariance} with technical supports.
\end{acknowledgments}

\bibliography{bnt_ia}

\end{document}